# Autonomous Uncertainty Quantification for Computational Point-of-care Sensors

Artem Goncharov[1], Rajesh Ghosh[2], Hyou-Arm Joung[1], Dino Di Carlo[2,3] and Aydogan Ozcan[1,2,3,4*]

[1]Electrical & Computer Engineering Department, [2]Bioengineering Department, [3]California NanoSystems Institute (CNSI), [4]Department of Surgery, University of California, Los Angeles, CA 90095 USA *Corresponding Author: ozcan@ucla.edu

## Abstract

Computational point-of-care (POC) sensors enable rapid, low-cost, and accessible diagnostics in emergency, remote and resource-limited areas that lack access to centralized medical facilities. These systems can utilize neural network-based algorithms to accurately infer a diagnosis from the signals generated by rapid diagnostic tests or sensors. However, neural network-based diagnostic models are subject to hallucinations and can produce erroneous predictions, posing a risk of misdiagnosis and inaccurate clinical decisions. To address this challenge, here we present an autonomous uncertainty quantification technique developed for POC diagnostics. As our testbed, we used a paper-based, computational vertical flow assay (xVFA) platform developed for rapid POC diagnosis of Lyme disease, the most prevalent tick-borne disease globally. The xVFA platform integrates a disposable paper-based assay, a hand-held optical reader and a neural network-based inference algorithm, providing rapid and cost-effective Lyme disease diagnostics in under 20 min using only 20 µL of patient serum. By incorporating a Monte Carlo dropout (MCDO)-based uncertainty quantification approach into the diagnostics pipeline, we identified and excluded erroneous predictions with high uncertainty, significantly improving sensitivity and reliability of the xVFA in an autonomous manner, without access to the ground truth diagnostic information of patients. Blinded testing using new patient samples demonstrated an increase in diagnostic sensitivity from 88.2% to 95.7%, indicating the effectiveness of MCDO-based uncertainty quantification in enhancing the robustness of neural network–driven computational POC sensing systems.

## Introduction

Point-of-care (POC) sensors represent an emerging class of diagnostic devices designed to enable rapid and inexpensive patient testing in emergency, underserved and remote settings that lack access to specialized equipment and highly trained medical personnel[1-3]. A common example of a POC sensor is a paper-based lateral flow assay (LFA)[4-6] – a disposal rapid diagnostic test that typically provides diagnostic results in <20 min using a single droplet of patient sample, such as whole blood[7], serum[8,9], saliva[10], or urine[11]. Conventional LFAs are primarily designed for single-analyte testing and can be interpreted visually by minimally trained or untrained users through the inspection of test and control lines within the active sensing region. In contrast, more advanced paper-based POC sensing systems have also emerged to support multiplexed testing[12,13], enabling the parallel detection of multiple biomarkers[14] or characterization of complex immune responses[15-17]. Such assays generate high-dimensional non-linear signals that are not readily interpretable by human users and therefore require integration with specialized readers and computational algorithms to process sensor outputs and infer diagnostic outcomes[3]. These computational POC sensors are particularly beneficial in scenarios that require clinician oversight, since test results generated at the POC can be remotely shared with physicians, who can then provide diagnostic interpretation and follow up clinical consultation.

Computational POC sensing platforms frequently employ advanced machine learning algorithms, such as neural networks, to accurately learn complex functional relationships between multiplexed sensor signals and underlying diagnostic states[14,15,18-21]. Despite their powerful function approximation capabilities, neural networks used in computational POC sensors are susceptible to hallucinations or unreliable predictions, which may result in erroneous diagnostic outcomes and compromise the reliability of computational biomedical sensing systems, potentially limiting their acceptance in clinical practice[3,22]. In addition, the "black-box" nature of neural networks reduces transparency of these algorithms, making it difficult to interpret the underlying decision-making process and potentially lowering trust among patients and clinicians[23,24]. To address these challenges, uncertainty quantification[25,26] techniques have been developed to estimate prediction uncertainty and improve the reliability of neural network-based diagnostics decision-making. In the context of computational POC sensing, uncertainty quantification can be applied to autonomously evaluate the confidence of individual diagnostic predictions and perform reliability testing that determines whether a given test outcome should be trusted or not – all without access to ground-truth patient information. As a result, each diagnostic outcome can be supplemented by a specific reliability action: "Trust" or "Do not use" (Figure 1). A "Trust" action indicates that the sensor has successfully passed the autonomous reliability assessment and the corresponding neural network prediction is considered trustworthy, whereas a "Do not use" action signifies that the prediction failed the reliability check and should be excluded from decision-making. Samples that fail this quality assurance check can undergo a follow up testing.

Two primary sources contribute to the uncertainty associated with neural network predictions: the first (epistemic uncertainty) arises from, e.g., limited or insufficient training data and/or model imperfections/bias, while the second (aleatoric uncertainty) originates from intrinsic noise and stochastic variability present within the measured data[27]. Both forms of uncertainty are inherently present in computational sensing systems: epistemic uncertainty, owing to the relatively small clinical datasets used to train computational POC sensors[28,29] and interpatient variability in immune responses that leads to different disease specific antibody titers; and aleatoric uncertainty, resulting from multiple sources of measurement noise and variations encountered in these systems (due to e.g., inexpensive paper materials, manual sensor operation, environmental factors, and sample matrix effects, etc.)[3,13,30]. Consequently, an effective uncertainty quantification strategy for computational POC sensing platforms should autonomously account for both epistemic and aleatoric uncertainties to ensure reliable diagnostic inference.

Monte Carlo dropout (MCDO) is a widely used Bayesian uncertainty quantification approach that applies stochastic dropout to the weights of a trained baseline neural network during the blind inference phase[31], addressing both epistemic and aleatoric uncertainties[32]. In MCDO, multiple forward passes with randomly sampled dropout masks are applied to each test sample during the blind testing stage, producing a distribution of predictions that can be compared against the baseline model output to quantify predictive uncertainty. Based on this uncertainty estimation, a reliability action – "Trust" or "Do not use" – can be assigned to the baseline prediction in an autonomous manner, without access to the ground truth information (Figure 2). The MCDO method has been successfully applied to quantify uncertainty and improve predictive performance in various computational diagnostic systems, including quantitative magnetic resonance imaging (MRI)[33], computed tomography (CT)[34], and optical coherence tomography (OCT)[35]. However, while MCDO demonstrated effectiveness in imaging-based systems, its applications to computational POC sensing platforms remain unexplored.

Here, we demonstrate MCDO-based uncertainty quantification for diagnostic quality assurance of computational POC sensing systems. As a testbed, we utilize a paper-based multiplexed vertical flow assay (xVFA) for the rapid and cost-effective diagnosis of Lyme disease (LD) [36,37], the most common tick-borne disease globally[38]. LD is a bacterial infection caused by *Borrelia burgdorferi* bacterium that is transmitted to humans through the bite of infected ticks (*Ixodes scapularis*)[39]. The current gold-standard laboratory diagnostics recommended by the US Centers for Disease Control and Prevention (CDC) rely on a tedious and costly two-tier testing procedure consisting of a first-tier enzyme immunoassay (EIA) followed by a second-tier western blot (WB) or a second EIA[40]. This multi-step testing strategy suffers from limited sensitivity during early-stage LD, when antibiotic treatment is most effective at preventing bacteria dissemination across the human body[41]. In addition, two-tier testing requires centralized laboratory infrastructure, leading to prolonged turnaround times and reduced accessibility to timely diagnostics, particularly in remote and rural areas where the LD prevalence is higher. To address these challenges of the two-tier testing process, a single-tier POC diagnostics platform for LD was developed based on the paper-based xVFA technology, enabling rapid and accessible LD diagnostics in POC settings[15,19,42]. This paper-based, cost-effective xVFA platform detects LD in under 20 min using a 20 µL droplet of patient serum. The assay incorporates a 2D sensing membrane functionalized with a curated panel of LD-specific peptides,[19,43-45] which selectively bind to human antibodies developed in response to the infection, enabling sensitive and specific serological detection. Following assay operation, an image of the activated sensing membrane is captured using a hand-held smartphone-based optical reader and processed by a neural network-based inference algorithm, in which a pre-trained network model analyzes multiplexed assay signals to infer LD diagnostic outcomes[15,19] (Figure 3a). To provide an autonomous quality assurance for the xVFA-based LD POC diagnostic platform, we developed an MCDO-based uncertainty quantification framework integrated into the diagnostic pipeline. In addition to processing each xVFA sample with the baseline diagnostic neural network, we performed inference using N = 50-1000 MCDO models, generating a distribution of prediction scores for each patient serum sample. By comparing the baseline prediction scores with the corresponding distribution of MCDO model scores, we introduced an uncertainty figure of merit (FOM) that autonomously quantifies diagnostic prediction uncertainty and excludes test samples with high uncertainty values from the decision-making process. Applying the MCDO-based uncertainty quantification pipeline to blinded patient samples improved LD diagnostic sensitivity from 88.2% to 95.7%, demonstrating the effectiveness of the proposed uncertainty quantification approach in enhancing the reliability of computational POC sensing systems. To the best of our knowledge, this study represents the first application of an MCDO-based uncertainty quantification technique to a computational POC sensing platform.

## Results and Discussion

### xVFA platform for Lyme disease diagnostics

The xVFA platform integrates a paper-based sensor, a hand-held optical reader, and a deep learning-based LD diagnostics algorithm, providing single-tier testing of LD in less than 20 min using a single droplet of patient serum. This cost-effective assay utilizes a vertical flow architecture that supports a 2D array of immunoreaction spots, enabling the incorporation of tens of individual reaction spots (channels) across the sensing membrane. These spots are functionalized with synthetic peptides containing *B. burgdorferi*-specific epitopes that efficiently bind human IgM and IgG antibodies developed in response to Lyme infection. In total, the LD sensing membrane contains 25 spots precoated with 9 distinct peptides in duplicate (i.e., 18 in total), along with 3 positive control spots (goat anti-mouse IgG) and 4 negative control spots (buffer). The assay operation takes ~16 min and incorporates two sequential steps: a sample loading step where human anti-borrelia IgM and IgG antibodies bind to the peptide-functionalized spots, and a signal generation step where gold nanoparticle (AuNP) conjugates react with human immunoglobulins, producing colorimetric optical signals (see the "Operation of xVFA for Lyme disease" section in the Methods). Following assay completion, an image of the activated sensing membrane is captured using the optical reader and colorimetric intensities from the immunoreaction spots are extracted from the image by a custom segmentation code executed on a computer. Finally, these extracted signals are processed by a neural network model to infer the LD diagnostic outcome for each patient (Figure 3a); see the Methods for details. The use of multiplexing in the xVFA design and the selected peptide panel is crucial for LD diagnostics since combining peptides that target multiple *B. burgdorferi*-specific epitopes reduces cross-reactivity with other bacterial infections while maintaining high diagnostic sensitivity for LD[19].

Before developing the autonomous uncertainty quantification framework to improve the reliability of neural network-based LD diagnostic inference, we first focused on selecting an appropriate baseline POC network model. Among various performance metrics for POC sensing systems, diagnostic sensitivity is often considered the most critical from the patient's health standpoint. False negative predictions, which lower the sensitivity of a test, leave infected individuals undetected and untreated, potentially leading to disease progression and associated complications, unless follow-up testing yields a true positive result. In the context of LD, the detection of earlier-stage infections is particularly beneficial due to the effectiveness of antibiotic-based treatments at these earlier time periods, whereas undiagnosed infections can progress to long-term health complications, including neurological, cardiac, or rheumatoid manifestations[46]. The negative implications of an increased false positive rate (which lowers specificity) for a POC test are less of a concern since additional testing beyond the initial POC screening would often be performed, which can mitigate false positive decisions. Therefore, to explicitly demonstrate the impact of uncertainty quantification on diagnostic performance, we intentionally focused on improving the sensitivity of the LD POC test in an autonomous manner, without access to ground-truth patient information. Accordingly, we selected a baseline neural network model, termed Lyme model ($L_0$), that utilized all the xVFA spots (including nine peptides) as input features to the network architecture; see Figure 3. The $L_0$ prediction scores range from 0 to 1, reflecting the output of the sigmoid activation function applied at the output layer of the LD diagnostic network; see the Methods section. The final diagnostic classification is performed by comparing the $L_0$ score to a score threshold of 0.5: samples with scores below 0.5 are classified as Lyme-negative, while samples with scores equal to or above 0.5 are classified as Lyme-positive.

The architecture of $L_0$ was selected based on its performance on a validation test set consisting of 93 activated xVFA cartridges from 31 patient serum samples obtained from the Lyme Disease Biobank (LDB). This selected model ($L_0$) demonstrated a specificity of 97.4% and a relatively low sensitivity of 81.5% on the validation dataset, as shown in Figure 4c (see "Baseline Lyme model architecture and training" section in Methods for more details on the architecture and training of $L_0$). The same model ($L_0$) was subsequently evaluated on an independent blind testing dataset consisting of 87 xVFA samples from 29 different patients collected from the same biobank, demonstrating a specificity of 100% and a sensitivity of 88.2% (see Figure 5c).

**Monte Carlo dropout (MCDO)-based autonomous uncertainty quantification for POC LD testing**

After optimizing the POC LD diagnostic model, $L_0$, next we developed an MCDO-based uncertainty quantification framework to improve sensitivity and overall reliability of the baseline LD diagnostic model in an autonomous manner, without access to patient ground truth information. In this pipeline, each xVFA sample under test is processed by N = 1000 MCDO Lyme models ($L_n$, n = 1…N), where each model applies a different random dropout mask with a 10% dropout rate to $L_0$ during the blind inference phase. Figures 4a and 5a show histograms of the prediction scores generated by these $L_n$ models (blue) overlaid with the corresponding $L_0$ outputs (green and pink) for the validation and blind testing datasets, respectively, with samples ordered from top to bottom by increasing $L_0$ scores (see "Generation of Monte Carlo dropout model histograms" section in Methods for more details). Similar to $L_0$ outputs, prediction scores generated by $L_n$ models also span the [0 : 1] range.

The distributions of $L_n$ scores are compared against the corresponding $L_0$ predictions on the validation dataset to define a quantitative uncertainty metric. By analyzing the absolute difference between the $L_0$ and the average $L_n$ prediction scores, we observed that the gap between these two sets of scores increases as the $L_0$ output approaches the decision threshold of 0.5. At the same time, the majority of false negative predictions in the validation dataset had $L_0$ scores clustered around the same threshold. Therefore, to quantitatively capture prediction uncertainty and potentially improve the predictive performance of the $L_0$ diagnostic model, we introduced a decision uncertainty figure of merit (F) as follows:

$$F = \frac{1}{|B_0 - <MC>|},$$

where $B_0$ is the $L_0$ prediction score (i.e., $B_0 = L_0(\mathbf{x})$) and <MC> is the average score of $L_n$ models defined as:

$$<MC> = \frac{1}{N}\sum_{n=1}^{N} L_n(\mathbf{x}),$$

$L_n(\mathbf{x})$ are the prediction scores of individual MCDO Lyme models (n = 1,…,N), and $\mathbf{x}$ is the array of xVFA input signals fed into the models.

Based on this definition, a low F value for a given patient sample indicates high predictive uncertainty, as the prediction score from $L_0$ and the average score generated by $L_n$ models significantly diverge. Accordingly, we introduced an uncertainty figure of merit threshold ($F_{th}$) as a cutoff to autonomously identify and exclude patient samples with low F values that are deemed unreliable due to elevated prediction uncertainty. Within this framework, uncertainty quantification for an xVFA sample is performed by computing its F value and comparing it to $F_{th}$. Each sample is then assigned one of the two reliability actions: "Trust" or "Do not use". Samples with F values greater than or equal to $F_{th}$ are classified as reliable ("Trust"), and the corresponding Lyme diagnostic predictions from $L_0$ are acceptable for clinical interpretation. In contrast, samples with F values less than $F_{th}$ are labeled as unreliable ("Do not use"), and the associated $L_0$ predictions are excluded from decision-making (Figure 3b,c).

The uncertainty threshold $F_{th}$ was optimized on the validation dataset by selecting a single cutoff value that removed a substantial fraction of false predictions at the expense of missing a minimal number of correctly classified samples. By using $F_{th} = 8.5$ and excluding samples with lower F values, we filtered out 7 of the 11 misclassified samples in the validation dataset at the expense of only 3 correctly classified true negative predictions (see Figure 4b and Figure S1a). This reliability testing step filtered out 6 false negatives and 1 false positive predictions, improving the validation test sensitivity of $L_0$ from 81.5% to 89.8% and increasing the overall validation accuracy from 88.2% to 94.0% (Figure 4c). The $L_0$ prediction scores for the excluded samples span a broad range between 0.24 and 0.59, indicating substantial variability in the baseline model output scores, and this variability limits the ability to assess model reliability from the prediction scores alone. In contrast, MCDO-based uncertainty quantification framework provides a quantitative metric (F) that efficiently filters out a significant portion of false predictions, irrespective of their baseline scores.

After developing the MCDO-based uncertainty quantification framework and optimizing $F_{th}$ on the validation dataset, we applied this pipeline to an independent blind testing cohort to evaluate its performance on previously unseen patient samples. Applying the $F_{th} = 8.5$ cutoff to the blind testing samples excluded 4 of the 6 false negative predictions, resulting in an increase in the blind testing sensitivity from 88.2% to 95.7% and an improvement in overall LD test accuracy from 94.2% to 97.4% (see Figure 5b,c and Figure S1b). These performance gains were achieved at the expense of excluding a small fraction of the correctly classified samples, specifically 1 true positive and 4 true negative predictions. In practice, all samples excluded during the reliability testing step can be directed for follow up testing to ensure reliable diagnostic outcomes for all patients. Although this process increases the sample-to-answer time for these cases, the quality assurance step helps minimize false negative diagnostic outputs, which could otherwise lead to severe long-term health complications due to undetected LD[46].

All the N = 1000 MCDO Lyme models used in the uncertainty quantification pipeline reported so far utilized a 10% dropout; however, additional dropout rates were evaluated for comparative analysis. For each dropout rate, the uncertainty threshold $F_{th}$ was independently optimized using the validation dataset. For dropout rates in the 1-50% range, we observed only minor variations in the optimal $F_{th}$ values as well as in the overall performance of the uncertainty quantification pipeline on the blind testing dataset (see Figures S2-S3). In contrast, larger dropout rates (≥50%) resulted in more significant variability, characterized by a broader distribution of F values for false negative samples and increased exclusion of true negative predictions (Figure S3b). This effect was particularly apparent at a 75% dropout rate, where a cutoff value of $F_{th} = 8.7$ excluded 11 true negative samples in addition to the 4 false negative predictions. Furthermore, a dropout rate of 90% caused a substantial shift of F values towards lower ranges for both true negative and false negative predictions. To partially address this behavior for larger dropout rates, we introduced two uncertainty thresholds, $F_{th}^{Low}$ and $F_{th}^{High}$, and excluded samples with F values falling between these boundaries (i.e., $F_{th}^{Low} \leq F \leq F_{th}^{High}$, see Figure S2c,f). Nevertheless, due to a significant mismatch in the F value distributions between the validation and blind testing datasets and the scattered distribution of F values for misclassified samples, the uncertainty quantification pipeline with a 90% dropout rate exhibited an inferior performance (see Figure S3a). Overall, these results indicate that lower dropout rates (≤20%) are preferable for MCDO-based uncertainty quantification in computational

POC sensing systems, as excessively large dropout percentages induce unstable discrepancies between the performances of $L_0$ and $L_n$.

Although we used N = 1000 $L_n$ models to generate a well-sampled distribution of prediction scores, we found that substantially smaller numbers of $L_n$ models can also achieve comparable performance on the blind testing dataset with a reduced computational cost. For example, using N = 50 $L_n$ models with the same uncertainty threshold ($F_{th}$ = 8.5) filtered out 5 false negative predictions in the validation set, and 4 false negative samples in the blind testing set, resulting in the same level of performance improvement in the blind testing sensitivity, achieving 95.7% (Figure S4). Generating N = 1000 MCDO predictions for 87 blind testing samples required 11.5 s of computation time, whereas the same process for N = 50 only took 0.66 s, reducing inference time by over an order of magnitude (Figure S5a). Nevertheless, both inference times are negligible compared to the total xVFA operation time (~16 min). In addition, the xVFA platform is primarily targeting decentralized and personalized LD testing, where testing batch sizes are typically small (on the order of 1-10 samples per run), further reducing inference times (i.e., to <10 s for N = 1000 and <0.6 s for N=50, Figure S5b,c). Therefore, a primary motivation for using a lower N in POC settings is not computational speed but rather practical considerations related to data storage and system constraints. In settings with limited data storage capacity, such as remote clinics lacking connectivity to centralized data centers or cloud computing services, data might be stored locally on benchtop computers or directly on the optical reader hardware, and lower N values may be beneficial to minimize data storage requirements and simplify data management.

To enhance the generalization capabilities of the uncertainty quantification framework, future efforts should expand the original training dataset to ideally include patient samples from multiple biobanks, various collection time periods, and diverse geographical regions. Such an expansion would enable the generation of a unified uncertainty threshold (or a set of thresholds) capable of supporting reliable and autonomous uncertainty quantification across diverse patient cohorts. As the xVFA platform and its associated neural network models are extended to a broader set of applications and target populations, the uncertainty quantification pipeline should be accordingly calibrated to ensure robust performance across all sample groups and diagnostic applications.

The MCDO-based uncertainty quantification framework demonstrated in this work requires access to the baseline Lyme model, $L_0$, to generate uncertainty estimates by using the dropout strategy. Consequently, implementation of MCDO-based reliability testing should be integrated into the diagnostic pipeline by the POC sensor developers/providers during the development stage, with a predetermined value of N. As the xVFA sensor and its associated neural network models evolve, the uncertainty quantification algorithm would need to be continuously updated and revalidated to ensure reliable performance across successive generations of POC sensors. As an alternative to this approach, model-free uncertainty quantification methods that enable reliability testing without direct access to the underlying diagnostic model are also important to consider, as they can facilitate the broader adoption of uncertainty quantification techniques by both sensor developers and end-users. Such model-free uncertainty quantification approaches would offer appealing alternatives to model-based methods, including the MCDO-based approach presented in this work, especially in scenarios where access to the baseline diagnostic models is limited or proprietary.

## Conclusions

In this work, we demonstrated an uncertainty quantification method based on Monte Carlo dropout to quantify diagnostic uncertainty and improve the performance of neural network-based computational POC sensing systems. As a proof-of-principle demonstration, we applied this approach to a multiplexed vertical flow assay (xVFA) platform for Lyme disease diagnostics, achieving an improvement in blind testing sensitivity up to 95.7%. To the best of our knowledge, this work represents the first application of MCDO-based uncertainty quantification techniques to a computational POC sensing platform. Providing an explicit uncertainty measure alongside diagnostic predictions is critical for improving trust and acceptance of computational POC sensors in clinical settings. Nevertheless, substantial work remains to be done before such systems can be routinely deployed in clinical practice and patient care. A key milestone in this transition is obtaining approvals from regulatory agencies, including the U.S. Food and Drug Administration (FDA), which is actively refining its regulatory framework to better accommodate machine learning–enabled medical devices[47]. In this context, uncertainty quantification and digital quality assurance techniques are expected to play an important role in these regulatory discussions since a robust and transparent uncertainty estimation framework can help improve model reliability, risk mitigation, and clinical safety, facilitating the regulatory review and approval processes. For example, samples that fail quality assurance step may undergo repeated testing to obtain a reliable diagnostic prediction, and in cases of repeated failure, when at least two consecutive test results are unreliable, the patient can be referred for a gold-standard laboratory-based diagnostics assay, such as a two-tier test for Lyme disease. This testing pipeline could ensure that patients receive trustworthy diagnostic results, while mitigating the risks associated with potential neural network model hallucinations. Overall, as computational POC sensing platforms continue to mature, close collaboration among sensor developers, regulatory bodies, and medical professionals will be crucial to facilitate the seamless integration of these devices into clinical workflows.

## Methods

### Operation of xVFA for Lyme disease

Prior to assay operation, all paper layers were prepared using a laser-cutter (Trotec) and a wax printer (Xerox) and stacked vertically within the top and bottom cases of the xVFA cartridge. Before initiating the assay, an image of the unused sensing membrane was captured using the optical reader to record the background signal. Then, the top and bottom cases were assembled using a twisting mechanism, and the assay was operated in two sequential stages. The first stage (sample loading), consisted of three steps: (i) loading 200 µL of running buffer to activate the assay, (ii) injection of a 20 µL serum sample, during which *B. burgdorferi*-specific antibodies bind to peptide-functionalized immunoreaction spots, and (iii) loading 200 µL of washing buffer to remove unbound antibodies and proteins from the sensing membrane. After an 8 min washing period, the first top case was removed and replaced with a second top case to initiate the signal generation stage. In this second stage, running buffer (200 µL), 40 nm AuNPs-detector antibody conjugates (a mixture of mouse anti-human IgM and IgG in a 1:1 ratio, 50 µL), and washing buffer (200 µL) were sequentially loaded, enabling visualization of anti-*Borrelia* antibodies through specific binding between human immunoglobulins and AuNP-conjugated detector antibodies. This step took an additional 8 min to complete, after which the second top case was opened, and an image of the activated sensing membrane was captured using the hand-held optical reader.

### Hand-held optical reader and image pre-processing

The optical reader used for xVFA readout consisted of a smartphone (LG G7 ThinQ) integrated with a custom 3D-printed optical attachment fabricated using Object 30 (Stratasys) and Ultimaker 3 (Ultimaker) 3D printers. This attachment was designed to securely hold the smartphone, an external optical module and xVFA cartridges to ensure stable and reproducible assay imaging. The optical module incorporated 4 green LEDs arranged in a circular pattern to provide uniform illumination of the sensing membrane, and an external lens to enable optimal focusing of the smartphone camera onto the sensing membrane plane. The LED peak emission wavelength (532 nm) was selected to match the peak absorption range of the AuNPs labels (520-560 nm) used for colorimetric signal generation. The LEDs were polished from the front to further improve illumination uniformity across the sensing membrane area. The 3D-printed attachment was also equipped with optical posts for pedestal installation, enabling both hand-held and benchtop operation. The xVFA sensing membrane imaging was performed using the standard LG smartphone camera application operated in manual mode, with imaging parameters set to ISO 50, an exposure time of 0.625 ms, and autofocus enabled.

For each assay, the optical reader captured images of the sensing membrane twice: before and after the assay was activated. Captured images were processed using a custom segmentation code on a computer to extract per-spot colorimetric signals by averaging pixel intensities within circular masks overlayed with the spots. Final absorption (colorimetric) signals from the 25 immunoreaction spots were calculated according to:

$$x_i^j = 1 - \frac{s_i^j}{b_i^j},$$

where $b_i^j$ and $s_i^j$ are the segmented per-spot signals before and after the xVFA activation, respectively, the index $i$ refers to the type of immunoreaction condition (9 peptide conditions, positive control, and negative control), and the index $j$ refers to the spot repeat within the given condition. These absorption signals (**x**) were used as inputs to the Lyme diagnostic models.

**Baseline Lyme model architecture and training**

The baseline Lyme diagnostic model ($L_0$) represents a shallow fully-connected binary classification neural network. The model used all 25 absorption signals (**x**) extracted from the xVFA sensing membrane as input features. The network architecture was selected through a hyperparameter grid search using the validation dataset of clinical samples. The selected LD diagnostic model included two hidden layers (256 and 64 units), each with ReLU activation functions and L2 regularization ($\lambda = 0.01$). Each hidden layer was followed by a batch normalization layer and a dropout layer with a dropout rate of 50% (which was only used during model training). The output layer contained 1 unit with a sigmoid activation function to generate a prediction score in the [0 : 1] range. The model was trained using a binary cross-entropy loss function, compiled with the Adam optimizer, a learning rate of 1e-2, and a batch size of 8. The binary cross-entropy loss is defined as:

$$L_{BCE} = -\frac{1}{N_b}\sum_{i=1}^{N_b}(y_i \log(y'_i) + (1 - y_i)\log(1 - y'_i)),$$

where $N_b$ is the batch size, $y_i$ are the ground truth labels ("0" for LD negative and "1" for LD positive), and $y'_i$ are the prediction scores ($y'_i \in [0 : 1]$) generated by the sigmoid activation function:

$$y'_i = \frac{1}{1 + e^{-\hat{y}_i}}$$

where $\hat{y}_i$ is the model's inference before the sigmoid function. The final diagnostics prediction per sample was determined by comparing the $y'_i$ score with the decision threshold of 0.5. Diagnostics predictions for samples with $y'_i < 0.5$ were assigned to Lyme-negative, while the predictions for the samples with $y'_i \geq 0.5$ were assigned to Lyme-positive.

The $L_0$ model was trained and validated on 93 xVFA samples activated with 31 unique patient serum samples. We utilized the same datasets for both training and validation due to the limited size of the training dataset. Employing conventional cross-validation approaches may lead to model overfitting on small validation subsets, leading to unreliable model performance on unseen data. Prior to the application of MCDO-based uncertainty quantification, $L_0$ achieved a sensitivity of 81.5% and a specificity of 97.4% on the validation dataset (Figure 4c). The $L_0$ model was further blindly tested on 87 xVFA samples from 29 different patients (never seen before), achieving a sensitivity of 88.2% and a specificity of 100%, as reported in the "Results and Discussion" section (Figure 5c). All the samples used for training, validation and blind testing were patient serum samples collected by the Lyme Disease Biobank (LDB).

To implement MCDO-based uncertainty quantification framework, each xVFA sample was first processed by the $L_0$ model without applying a dropout during the blind inference stage. At this step, the effective dropout rate was 0%, since the 50% dropout layers specified in the $L_0$ architecture were only active during model training. This step generated the baseline score ($y_i'$). Subsequently, each sample was processed by N = 50-1000 MCDO Lyme models that shared the same architecture as $L_0$, but had random dropout masks applied at a 10% dropout rate during the blind inference stage, generating a distribution of MCDO scores ($L_n(\mathbf{x}), n = 1, \ldots, N$). The average MCDO score and the corresponding uncertainty figure of merit (F) were then computed according to the equations defined in the "Results and Discussion" section.

All neural network models were trained and evaluated on a desktop computer equipped with a GeForce GTX 1080 Ti graphics processing unit (NVIDIA). The MCDO-based uncertainty quantification pipeline was developed in Python using NumPy, TensorFlow, and Keras libraries. Training time for $L_0$ was ~20 s, while the blind testing of the trained $L_0$ model was substantially faster, taking less than 0.3 s to process all 87 blind testing samples. Blind testing using N = 1000 MCDO Lyme models required 11.5 s, which was reduced to 0.66 s when using N = 50 MCDO Lyme models (Figure S5b,c).

**Generation of Monte Carlo dropout model histograms**

Figures 4a and 5a contain histograms of $L_n$ scores for the validation and blind testing datasets, respectively. To generate these histograms, each xVFA sample was first processed by N=1000 $L_n$ models, each producing a prediction score in the [0 : 1] range based on the sigmoid activation function at the network output layer. Next, these scores were discretized into 50 bins, each with a width of 0.02, generating a 1D histogram for every sample. In this representation, the value of each bin represents $L_n$ models density calculated as:

$$\rho_j = \frac{1}{N} \sum_{n=1}^{N} rect(\frac{L_n(\mathbf{x})}{w} - j - \frac{1}{2}) * 100\%,$$

where $j = 0\ldots49$ is the bin number and $w = 0.02$ is the bin width. Based on this definition, $\rho_j$ takes values between 0% and 100%; $\rho_j = 0\%$ if no $L_n$ models for this sample have output scores within bin j and $\rho_j = 100\%$ if all N=1000 $L_n$ models for a given sample have scores within bin j.

Each 1D histogram was further overlaid with the corresponding $L_0$ score classified into one of the 50 bins. For each patient sample, the $L_0$ score ($B_0$) was assigned to a bin number $\lfloor B_0/w \rfloor$, and the corresponding bin was marked with a green (for Lyme negative predictions) or a pink (for Lyme positive predictions) circle. The resulting sample-wise 1D histograms were then stacked row-wise to form a 2D histogram, in which each row represented an individual test sample and each column corresponded to a different classification bin, covering the $L_n$ score distribution in the [0 : 1] range. Samples in this 2D histogram were sorted from top to bottom according to increasing $L_0$ scores.

# Supporting Information:

Supplementary Figures S1-S5.

# References

[1] Kaushik, A. & Mujawar, M. A. Point of care sensing devices: better care for everyone. *Sensors* **18**, 4303 (2018);
[2] Zhang, W., Wang, R., Luo, F., Wang, P. & Lin, Z. Miniaturized electrochemical sensors and their point-of-care applications. *Chin. Chem. Lett.* **31**, 589-600 (2020);
[3] Han, G. R., Goncharov, A., Eryilmaz, M., Ye, S., Palanisamy, B., et al. Machine learning in point-of-care testing: Innovations, challenges, and opportunities. *Nat. Commun.* **16**, 3165 (2025);
[4] Budd, J., Miller, B. S., Weckman, N. E., Cherkaoui, D., Huang, D., et al. Lateral flow test engineering and lessons learned from COVID-19. *Nat. Rev. Bioeng.* **1**, 13-31 (2023);
[5] Kakkar, S., Gupta, P., Yadav, S. P. S., Raj, D., Singh, G., et al. Lateral flow assays: Progress and evolution of recent trends in point-of-care applications. *Mater. Today Bio* **28**, 101188 (2024);
[6] Kinyua, D. M., Memeu, D. M., Mugo Mwenda, C. N., Ventura, B. D., & Velotta, R. Advancements and applications of lateral flow assays (LFAs): A comprehensive review. *Sensors* **25**, 5414 (2025);
[7] Jung, Y., Kim, S., Kim, M. G., Lee, Y. E., Shin, M. G., & Yang, S. One-step detection of vancomycin in whole blood using the lateral flow immunoassay. *Biosensors* **14**, 129 (2024);
[8] Han, G. R., & Kim, M. G. Highly sensitive chemiluminescence-based lateral flow immunoassay for cardiac troponin I detection in human serum. *Sensors* **20**, 2593 (2020);
[9] Kim, H. T., Jin, E., & Lee, M. H. Portable chemiluminescence-based lateral flow assay platform for the detection of cortisol in human serum. *Biosensors* **11**, 191 (2021);
[10] Jung, C., & Kim, M. G. Direct use of a saliva-collected cotton swab in lateral flow immunoassay for the detection of cotinine. *Biosensors* **12**, 214 (2022);
[11] Patil, A. A., Kaushik, P., Jain, R. D., & Dandekar, P. P. Assessment of urinary biomarkers for infectious diseases using lateral flow assays: a comprehensive overview. *ACS Infect. Dis.* **9**, 9-22 (2022);
[12] Ballard, Z., Brown, C., Madni, A. M., & Ozcan, A. Machine learning and computation-enabled intelligent sensor design. *Nat. Mach. Intell.* **3**, 556-565 (2021);
[13] Jeon, J., Choi, H., Han, G. R., Ghosh, R., Palanisamy, B., Di Carlo, D., Ozcan, A., & Park, S. Paper-based Vertical Flow Assays for in Vitro Diagnostics and Environmental Monitoring. *ACS sensors* **10**, 3317-3339 (2025);
[14] Goncharov, A., Joung, H. A., Ghosh, R., Han, G. R., Ballard, Z. S., et al. Deep Learning-Enabled Multiplexed Point-of-Care Sensor using a Paper-Based Fluorescence Vertical Flow Assay. *Small* **19**, 2300617 (2023);
[15] Joung, H. A., Ballard, Z. S., Wu, J., Tseng, D. K., Teshome, H., et al. Point-of-care serodiagnostic test for early-stage Lyme disease using a multiplexed paper-based immunoassay and machine learning. *ACS nano* **14**, 229-240 (2019);
[16] Zhang, T., Deng, R., Wang, Y., Wu, C., Zhang, K., et al. A paper-based assay for the colorimetric detection of SARS-CoV-2 variants at single-nucleotide resolution. *Nat. Biomed. Eng.* **6**, 957-967 (2022);
[17] Eryilmaz, M., Goncharov, A., Han, G. R., Joung, H. A., Ballard, Z. S., et al. A paper-based multiplexed serological test to monitor immunity against SARS-COV-2 Using machine learning. *ACS nano* **18**, 16819-16831 (2024);

[18] Ballard, Z. S., Joung, H. A., Goncharov, A., Liang, J., Nugroho, K., Di Carlo, D., Garner, O. B., & Ozcan, A. Deep learning-enabled point-of-care sensing using multiplexed paper-based sensors. *NPJ digit. Med.* **3**, 66 (2020);

[19] Ghosh, R., Joung, H. A., Goncharov, A., Palanisamy, B., Ngo, K., et al. Rapid single-tier serodiagnosis of Lyme disease. *Nat. commun.* **15**, 7124 (2024);

[20] Han, G. R., Goncharov, A., Eryilmaz, M., Joung, H. A., Ghosh, R., et al. Deep learning-enhanced paper-based vertical flow assay for high-sensitivity troponin detection using nanoparticle amplification. *ACS nano* **18**, 27933-27948 (2024);

[21] Han, G. R., Goncharov, A., Eryilmaz, M., Ye, S., Joung, H. A., Deep Learning-Enhanced Chemiluminescence Vertical Flow Assay for High-Sensitivity Cardiac Troponin I Testing. *Small* **21**, 2411585 (2025);

[22] Hatem, R., Simmons, B., & Thornton, J. E. A call to address AI "hallucinations" and how healthcare professionals can mitigate their risks. *Cureus* **15**, (2023);

[23] Zhang, J., & Zhang, Z. M. Ethics and governance of trustworthy medical artificial intelligence. *BMC Med. Inform. Decis. Mak.* **23**, 7 (2023);

[24] Wadden, J. J. Defining the undefinable: the black box problem in healthcare artificial intelligence. *J. Med. Ethics* **48**, 764-768 (2022);

[25] Lambert, B., Forbes, F., Doyle, S., Dehaene, H., & Dojat, M. Trustworthy clinical AI solutions: A unified review of uncertainty quantification in Deep Learning models for medical image analysis. *Artif. Intell. Med.* **150**, 102830 (2024).

[26] Abdar, M., Khosravi, A., Islam, S. M. S., Acharya, U. R., & Vasilakos, A. V. The need for quantification of uncertainty in artificial intelligence for clinical data analysis: increasing the level of trust in the decision-making process. *IEEE SMC* **8**, 28-40 (2022);

[27] Hüllermeier, E., & Waegeman, W. Aleatoric and epistemic uncertainty in machine learning: An introduction to concepts and methods. *Mach. Learn.* **110**, 457-506 (2021);

[28] Evans, H., & Snead, D. Why do errors arise in artificial intelligence diagnostic tools in histopathology and how can we minimize them?. *Histopathol.* **84**, 279-287 (2024);

[29] Celi, L. A., Cellini, J., Charpignon, M. L., Dee, E. C., Dernoncourt, F. Sources of bias in artificial intelligence that perpetuate healthcare disparities—A global review. *PLOS digit. health* **1**, e0000022 (2022);

[30] Flynn, C. D., & Chang, D. Artificial intelligence in point-of-care biosensing: challenges and opportunities. *Diagnostics* **14**, 1100 (2024);

[31] Gal, Y., & Ghahramani, Z. Dropout as a bayesian approximation: Representing model uncertainty in deep learning. *PMLR*, 1050-1059 (2016);

[32] Kwon, Y., Won, J. H., Kim, B. J., & Paik, M. C. Uncertainty quantification using Bayesian neural networks in classification: Application to biomedical image segmentation. *Comput. Stat. Data Anal.* **142**, 106816 (2020);

[33] Avci, M. Y., et al. Improving accuracy and uncertainty quantification of deep learning based quantitative MRI using Monte Carlo dropout. arXiv Preprint at (2021);

[34] Ahn, S. H., et al. Uncertainty Quantification in Automated Detection of Vertebral Metastasis Using Ensemble Monte Carlo Dropout. *JIIM*, 1-16 (2024);

[35] Orlando, J. I., et al. U2-net: A bayesian u-net model with epistemic uncertainty feedback for photoreceptor layer segmentation in pathological oct scans. *ISBI*, 1441-1445 (2019);

[36] Steere, A. C. Lyme disease. *N. Engl. J. Med.* **345**, 115–125 (2001);

[37] Bobe, J. R., et al. Recent progress in Lyme disease and remaining challenges. *Front. Med.* **8**, 666554 (2021);

[38] Stanek, G., Wormser, G. P., Gray, J., & Strle, F. Lyme borreliosis. *The Lancet* **379**, 461-473 (2012);

[39] Rosenberg, R. Vital signs: trends in reported vectorborne disease cases—United States and Territories, 2004–2016. *MMWR* **67**, (2018);

[40] Moore, A., Nelson, C., Molins, C., Mead, P. & Schriefer, M. Current guidelines, common clinical pitfalls, and future directions for laboratory diagnosis of Lyme disease, United States. *Emerg. Infect. Dis.* **22**, 1169–1177 (2016);

[41] Branda, J. A. et al. Advances in serodiagnostic testing for Lyme disease are at hand. *Clin. Infect. Dis.* **66**, 1133–1139 (2018);

[42] Joung, H.-A. et al. Point-of-care serodiagnostic test for early-stage Lyme disease using a multiplexed paper-based immunoassay and machine learning. *ACS Nano* **14**, 229–240 (2020);

[43] Signorino, G., Arnaboldi, P. M., Petzke, M. M. & Dattwyler, R. J. Identification of OppA2 linear epitopes as serodiagnostic markers for Lyme disease. *Clin. Vaccin. Immunol.* **21**, 704–711 (2014);

[44] Toumanios, C., Prisco, L., Dattwyler, R. J. & Arnaboldi, P. M. Linear B cell epitopes derived from the multifunctional surface lipoprotein BBK32 as targets for the serodiagnosis of Lyme disease. *mSphere* **4**, e00111–e00119 (2019);

[45] Arnaboldi, P. M., Katseff, A. S., Sambir, M. & Dattwyler, R. J. Linear peptide epitopes derived from ErpP, p35, and FlaB in the serodiagnosis of Lyme disease. *Pathogens* **11**, 944 (2022);

[46] Wormser, G. P. et al. The clinical assessment, treatment, and prevention of Lyme disease, human granulocytic anaplasmosis, and Babesiosis: clinical practice guidelines by the Infectious Diseases Society of America. *Clin. Infect. Dis.* **43**, 1089–1134 (2006).

[47] Proposed Regulatory Framework for Modifications to Artificial Intelligence/Machine Learning (AI/ML)-Based Software as a Medical Device (SaMD), U.S. Food & Drug Administration webpage, https://www.fda.gov/files/medical%20devices/published/US-FDA-Artificial-Intelligence-and-Machine-Learning-Discussion-Paper.pdf (2019).

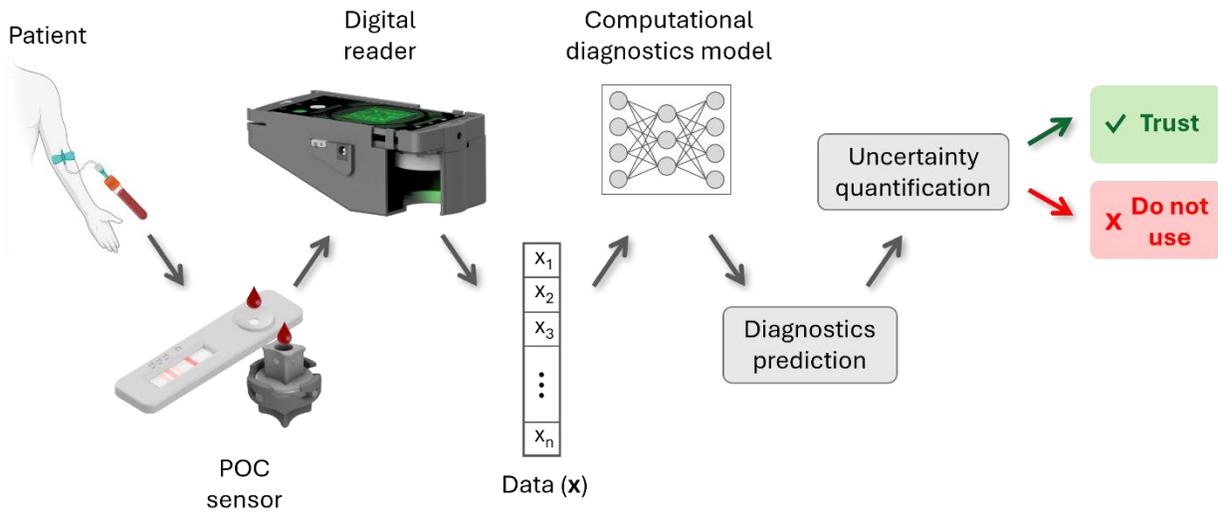

**Figure 1.** Overview of the computational point-of-care sensing pipeline with uncertainty quantification. The pipeline outputs a reliability action for each sensor, classified as either "Trust" or "Do not use". A "Trust" outcome indicates that the sensor measurement has passed reliability testing and that the corresponding diagnostic prediction can be used in the clinical decision-making process. In contrast, a "Do not use" outcome signifies that the test result is unreliable and that the associated diagnostic prediction should be excluded from the diagnostic workflow.

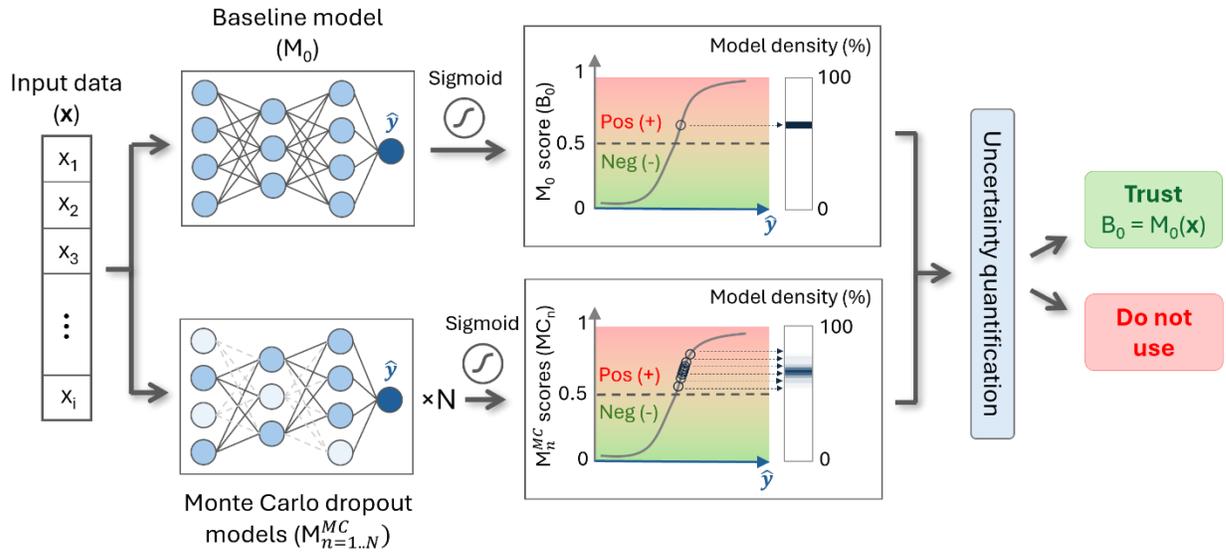

**Figure 2.** Monte Carlo dropout (MCDO)-based uncertainty quantification method for a classification neural network model. The baseline model score is compared with the distribution of scores generated by N MCDO models to perform uncertainty quantification in an autonomous manner, without access to ground truth patient information. Based on this comparison, each sample undergoes reliability assessment and is assigned one of the two reliability actions: "Trust", indicating that the prediction has passed the uncertainty testing, or "Do not use", indicating that the prediction has failed the test and should be excluded from the clinical workflow.

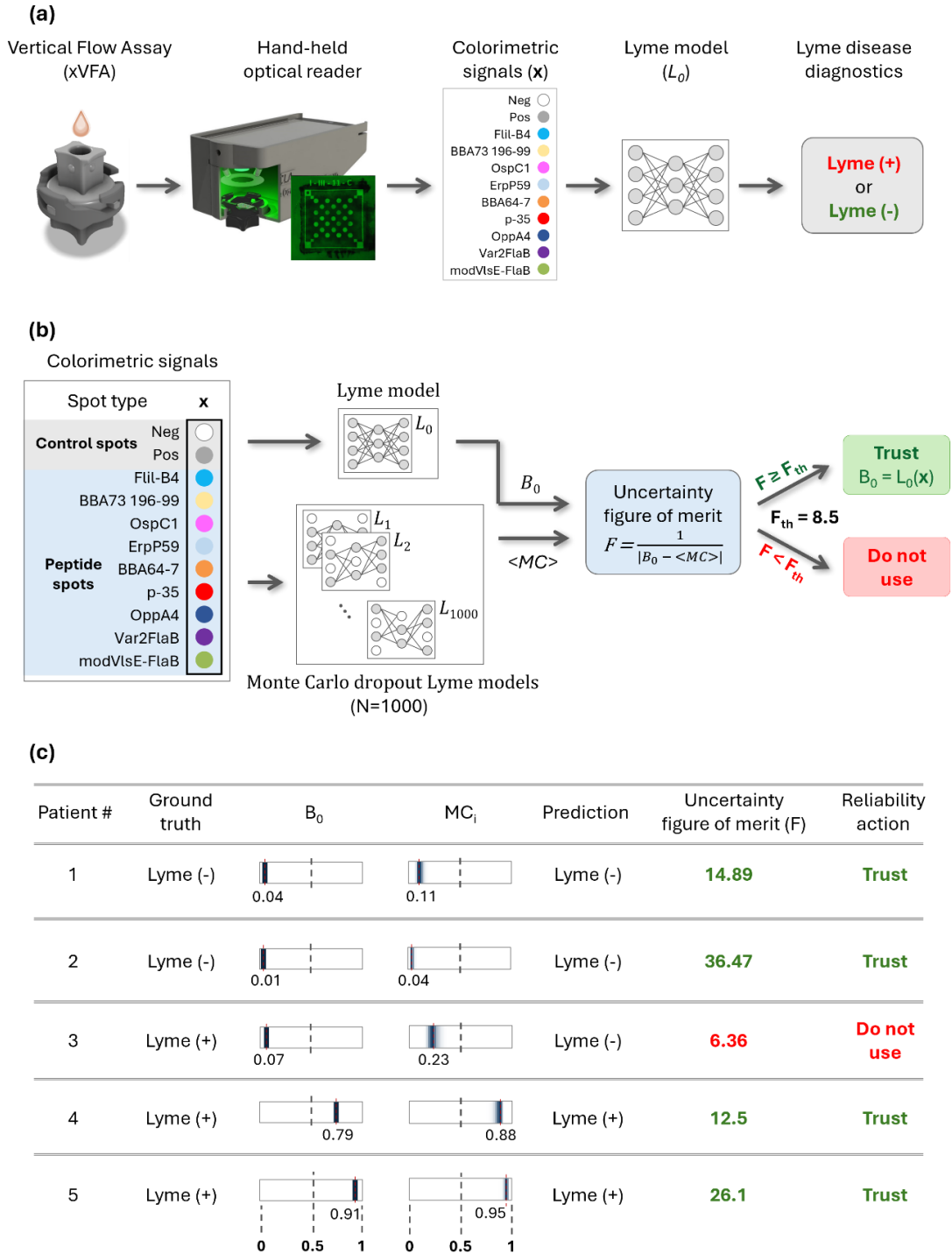

**Figure 3.** (a) Overview of the multiplexed vertical flow assay (xVFA) platform for Lyme disease diagnostics. (b) Monte Carlo dropout (MCDO)-based uncertainty quantification pipeline in the xVFA platform. For each xVFA sample, the score from the Lyme diagnostic model ($L_0$) is compared with the score distribution from the MCDO Lyme models ($L_n$). Based on this comparison, an uncertainty figure of merit (F) is computed and evaluated against the uncertainty figure of merit threshold ($F_{th}$) to assign reliability action to each samples as either "Trust" or "Do not use". Samples with $F \geq F_{th}$ are labeled as "Trust", indicating that the corresponding diagnostic predictions can be used for clinical decision-making. Samples with $F < F_{th}$ are labeled as "Do not use", indicating unreliable predictions that should be excluded from diagnostic interpretation. (c) Table for a few representative patient samples from the blind testing set summarizing ground truth Lyme diagnostics, $L_0$ scores and diagnostic predictions, distributions of $L_n$ scores, computed F values, and the resulting reliability actions.

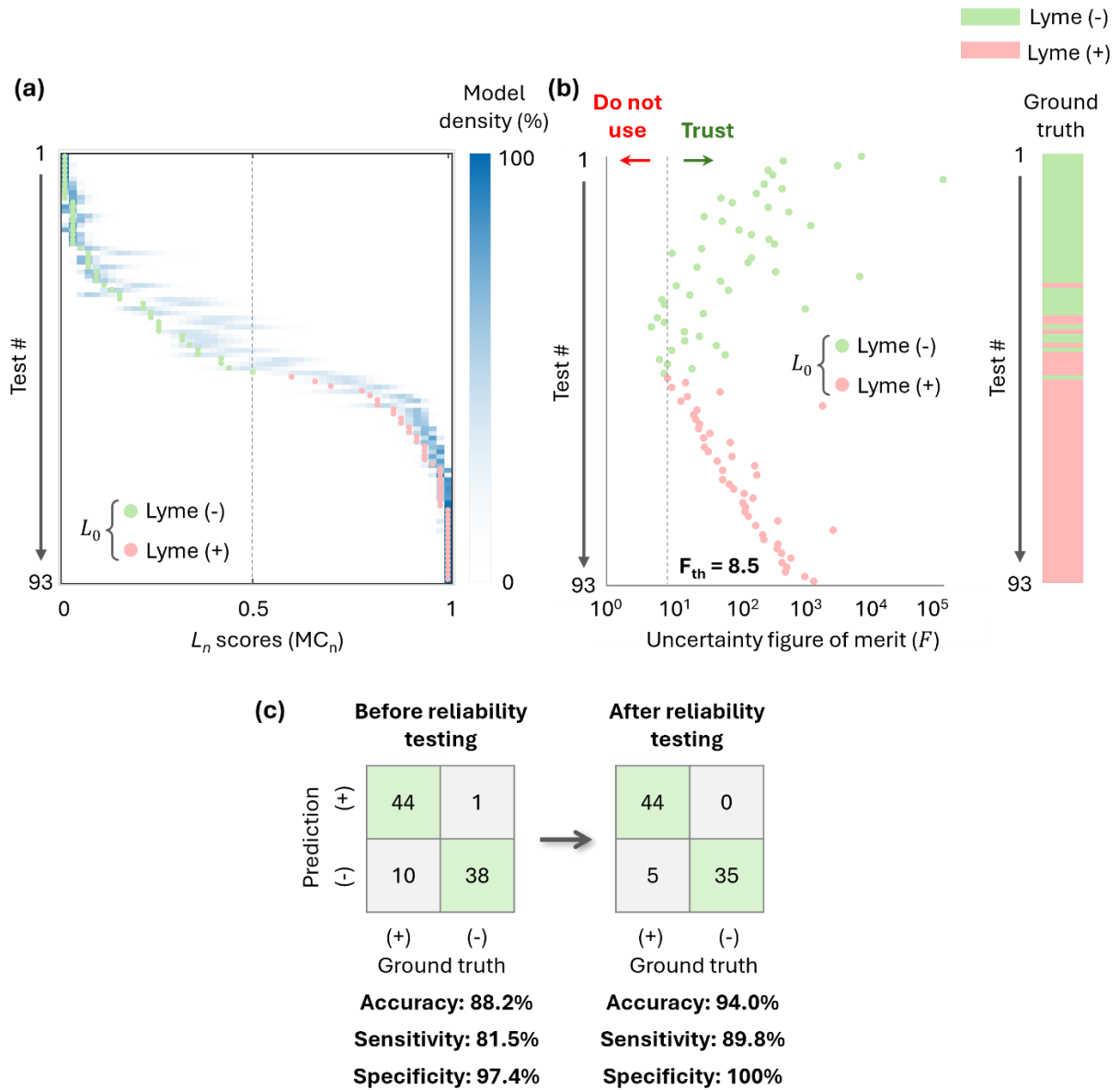

**Figure 4.** (a) Histogram, summarizing the Lyme model ($L_0$) scores and the corresponding distributions of the MCDO-based Lyme model ($L_n$) outputs for 93 xVFA samples from the validation dataset, sorted from top to bottom by increasing $L_0$ scores. (b) Uncertainty figure of merit (F) computed for samples from the validation dataset. Green and pink colors represent Lyme negative and Lyme positive samples, respectively, as predicted by $L_0$. Each sample is assigned a reliability action based on its F value. Samples with F values below uncertainty figure of merit threshold ($F_{th}$ = 8.5) are assigned a "Do not use" action, indicating that the corresponding $L_0$ predictions are unreliable, and should be excluded from clinical decision-making. Samples with F values above or equal to $F_{th}$ are assigned a "Trust" action, indicating that the corresponding $L_0$ predictions are reliable and suitable for clinical interpretation. (c) $L_0$ diagnostic predictions for the validation dataset before (left) and after (right) the application of the MCDO-based reliability testing step.

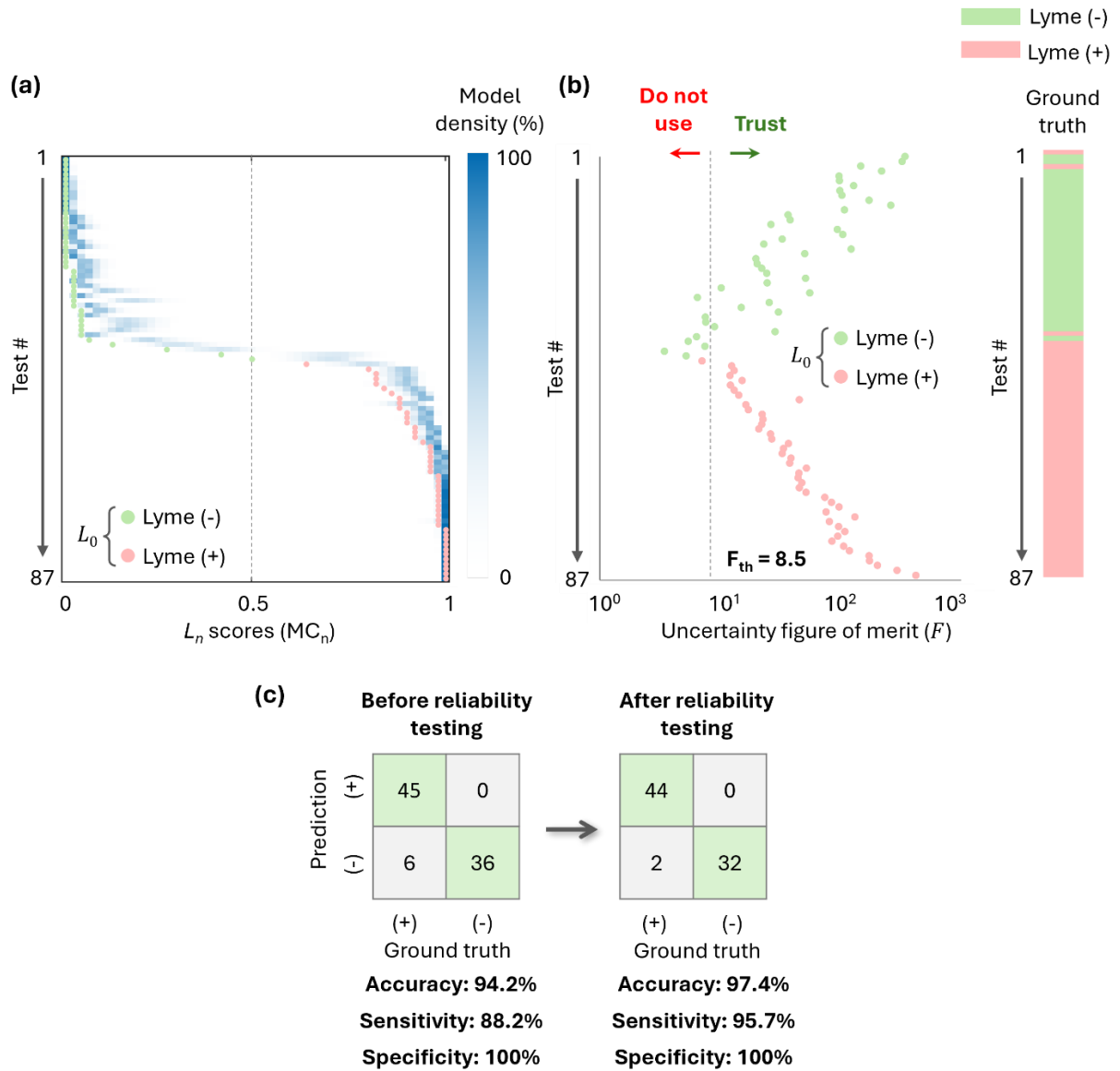

**Figure 5.** Same as Figure 4, except for the blind testing dataset with new patient samples, never seen before. (a) Histogram, summarizing the Lyme model ($L_0$) scores and the corresponding distributions of Monte Carlo dropout (MCDO) Lyme model ($L_n$) outputs for 87 xVFA samples from the blind testing dataset, sorted from top to bottom by increasing $L_0$ scores. (b) Uncertainty figure of merit (F) computed for samples from the blind testing dataset. Green and pink colors represent Lyme negative and Lyme positive samples, respectively, as predicted by $L_0$. Each sample is assigned a reliability action based on its F value. Samples with F values below uncertainty figure of merit threshold ($F_{th}$ = 8.5) are assigned a "Do not use" action, indicating that the corresponding $L_0$ predictions are unreliable, and should be excluded from clinical decision-making. Samples with F values above or equal to $F_{th}$ are assigned a "Trust" action, indicating that the corresponding $L_0$ predictions are reliable and suitable for clinical interpretation. (c) $L_0$ diagnostic predictions for the blind testing dataset before (left) and after (right) the application of the MCDO-based reliability testing step.

18